\documentclass{appolb}
\usepackage{graphicx}

\usepackage{lineno}

\begin{document}
\title{Exclusive Jet Measurement in Special LHC Runs\\Feasibility Studies %
\thanks{Presented at XXII Cracow Epiphany Conference}
\author{Maciej Trzebi\'nski
\thanks{This work was supported in part by Polish Ministry of Science and Higher Education under the Mobility Plus programme (1285/MOB/IV/2015/0).}%
}
\address{Institute of Nuclear Physics Polish Academy of Sciences\\
152 Radzikowskiego St., Krak ́ow, Poland}
}
\maketitle
\begin{abstract}

The feasibility studies of the central exclusive jet production at the LHC using the proton tagging technique are presented. Three classes of data taking scenarios are considered: double tag at high pile-up, single tag at low pile-up and double tag at low pile-up. Analyses were performed at the c.m. energy of 14 TeV for the ATLAS experiment, but are also valid for the CMS/TOTEM detectors.

\end{abstract}
  
\section{Introduction}
\begin{figure}
  \centering
  \includegraphics[width=0.22\columnwidth]{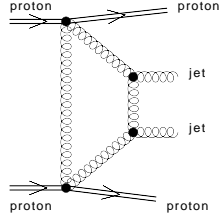}
\caption{Central exclusive jet production: both interacting protons stay intact and two jets are produced.}
\label{fig_exc_jj_diag}
\end{figure}

The central exclusive jet events constitute a special class among diffractive jet productions. In such events both protons stay intact and all energy is used to produce central system, see Figure~\ref{fig_exc_jj_diag}. Hence, the kinematic properties of scattered protons are connected to the ones of produced central system.

There are several theoretical descriptions of the exclusive jet production mechanisms. In this paper, the model of Khoze-Martin-Ryskin (KMR)~\cite{KMR} is used. In the KMR model a perturbative approach is applied -- the colourless exchange is represented by an exchange of two gluons: a hard and a soft one. The role of the soft gluon is to provide the colour screening that ensures that no net colour charge is exchanged between the two protons.

In order to perform a fully exclusive measurement both the jets and the intact protons need to be measured. The requirement of both protons being tagged often implies the production of a large amount of energy in the central region. In consequence, the cross-section is significantly reduced and a large amount of data need to be collected (high pile-up environment). Such measurement was shown to be feasible~\cite{ATLAS_exclusive} and will be discussed in details in the next section. Since fully exclusive measurement in a high pile-up environment is very challenging, in~\cite{EXC_JJ_one_tag} a semi-exclusive approach was proposed and shown to be feasible. Finally, for the first time, in this publication, a low-luminosity measurement with both protons tagged is discussed.

\section{Experimental Environment}
The exclusive production can be characterised by the presence of two jets in the central region and two scattered protons. Jets are assumed to be measurable by central ATLAS detector~\cite{ATLAS} whereas protons need to be tagged by dedicated devices -- so called forward detectors. At present, ATLAS is equipped with two sets of such apparatus: ALFA~\cite{ALFA} and AFP~\cite{AFP}.

Diffractive protons are scattered at very low angles. Due to this, proton detectors are located far away from the Interaction Point (IP): 205 and 217~m in case of AFP and 237 and 245 m in case of ALFA. Moreover, proton trajectory between IP and location of forward detector is not a straight line. This is due to the presence of LHC magnets. The settings of these magnets, commonly called \textit{optics}, play a key role in the exclusive analysis. In the simplest possible way these settings could be characterized by the value of the betatron function at the IP, $\beta^*$. Studies presented in this paper focuses on two such settings: $\beta^* = 0.55$~m and $\beta^* = 90$~m. The detailed description of the properties of these optics sets can be found in~\cite{LHC_optics} whereas the justification of the choice is explained in~\cite{EXC_JJ_one_tag}.

\section{High Luminosity, Double Tagged Measurement}
Feasibility studies of the exclusive jet production with forward proton tagging were performed for the ATLAS experiment, AFP detectors and $\beta^* = 0.55$~m optics~\cite{ATLAS_exclusive}. They were followed by similar ones done by CMS/TOTEM~\cite{yellow_report}. Due to the limited acceptance, the lowest jet $p_T$ was of about 150 GeV. This requires using high integrated luminosity, thus high pile-up. In the analysis two scenarios were considered: $\mu = 23$ with integrated luminosity of 40 fb$^{-1}$ and $\mu = 46$ with $L = 300$ fb$^{-1}$. After the selection, a signal-to-background ratio was found to be of 0.57 (0.16) for $\mu$ = 23 (46).

For both considered scenarios the statistical errors were found to be considerably small. The biggest uncertainty was associated with the modelling of the combinatorial background from non-diffractive dijet events overlapped with two protons from pile-up events.

\section{Low Luminosity, Single Tagged Measurement}
The drawback of studies described above was the need to collect data in a harsh, high pile-up environment. To address this issue, it was proposed to perform a measurement in a semi-exclusive mode, when only one of the scattered protons is tagged~\cite{EXC_JJ_one_tag}. This lowered the acceptance limits towards jets with smaller $p_T$ values.

In the analysis, performed at $\sqrt{s} = 14$ TeV, four data-taking scenarios were considered: AFP and ALFA detectors as forward proton taggers and $\beta^* = 0.55$ m, $\beta^* = 90$ m optics. After the signal selection, the signal-to-background ratio was found to between 5 and 10~000, depending on the considered run scenario. Moreover, it was shown that the significant measurement can be carried out for the data collection period of about 100~h with pile-up multiplicity of $\mu \sim 1$.

\section{Low Luminosity, Double Tagged Measurement}
To complete the studies described above, int this paper a double tagged, low-$p_T$ exclusive jet measurement is discussed. It can be carried out with ALFA detectors and $\beta^* = 90$ m optics.

\subsection{Monte Carlo Generators and Event Reconstruction}
Exclusive jets were generated accordingly to KMR model~\cite{KMR} using \textsc{FPMC}~\cite{FPMC}. This event generator is a modification of \textsc{Herwig 6.5}~\cite{Herwig} and uses its final state parton showering and hadronisation algorithms.

Diffractive jet backgrounds (double Pomeron exchange (DPE) and single diffractive (SD)) were also generated using \textsc{FPMC}, assuming the rapidity gap survival factor of 0.03 and 0.1, respectively~\cite{gap_surv}. The generation was based on the the resolved Pomeron model~\cite{Ingelman} and H1 2007 Fit B~\cite{Hera_fit}. The multi-parton interactions (MPI) were turned off. The non-diffractive (ND) jets and minimum-bias events were generated using \textsc{Pythia8}~\cite{Pythia8}.

Jets were reconstructed using the anti-$k_T$ algorithm implemented in the FastJet package~\cite{FastJet} with the jet radius $R = 0.6$. Scattered protons were transported to the location of the considered forward detector by the \textsc{FPTrack} program~\cite{FPTrack}. The proton energy was reconstructed using the procedure described in~\cite{unfolding}.

\subsection{Signal Selection}
Due to ATLAS jet reconstruction performance, the minimal jet transverse momentum was required to be greater than 20 GeV. Event was accepted, when it contained at least two reconstructed jets and two protons detected in ALFA stations. The distance between the beam centre and the detector active area was set to 6.9 mm, which corresponds to 10 nominal beam widths plus 0.3 mm of detector dead material.

\begin{figure}
  \centering
  \hfill
  \includegraphics[width=0.42\columnwidth]{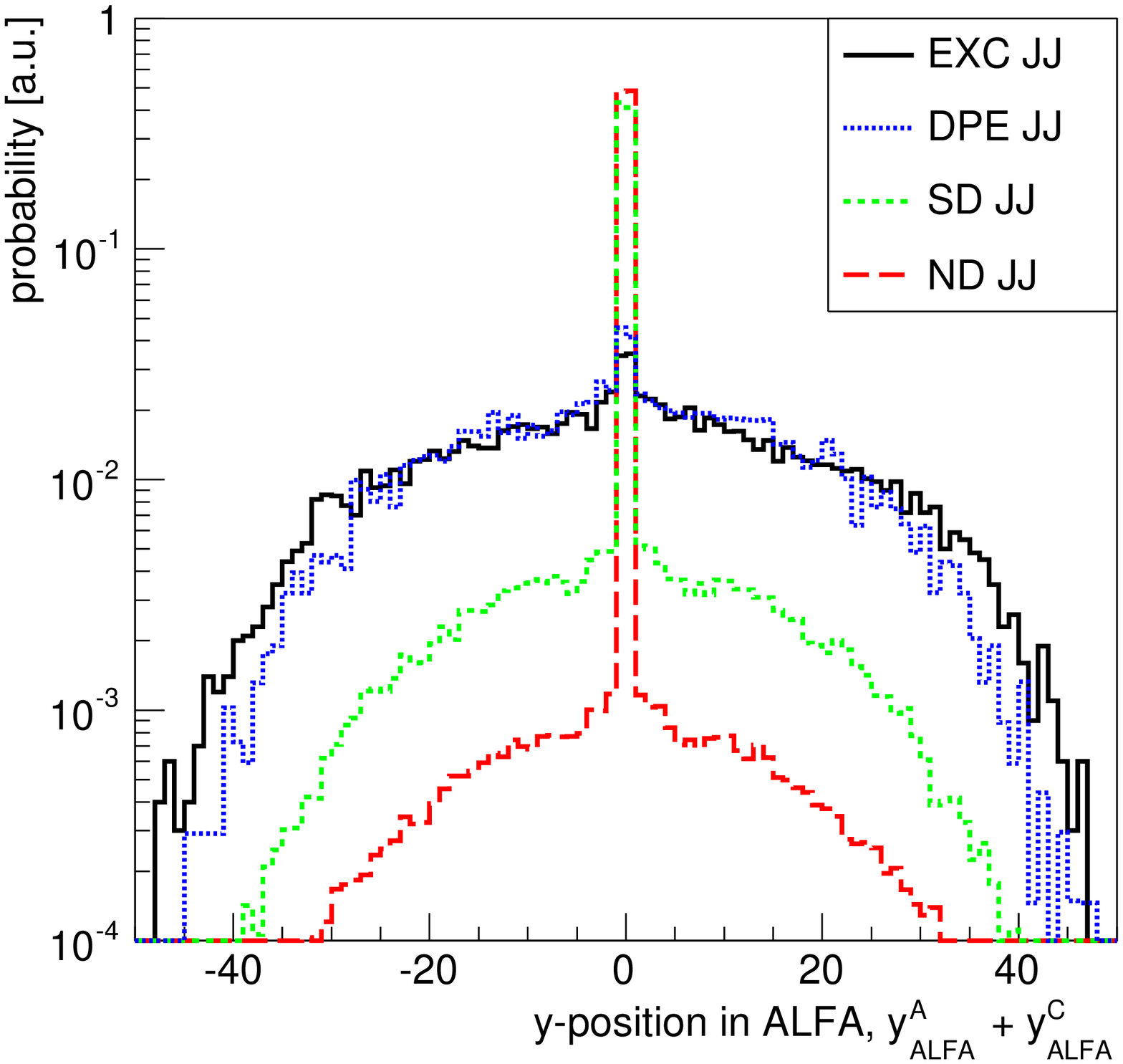}\hfill
  \includegraphics[width=0.42\columnwidth]{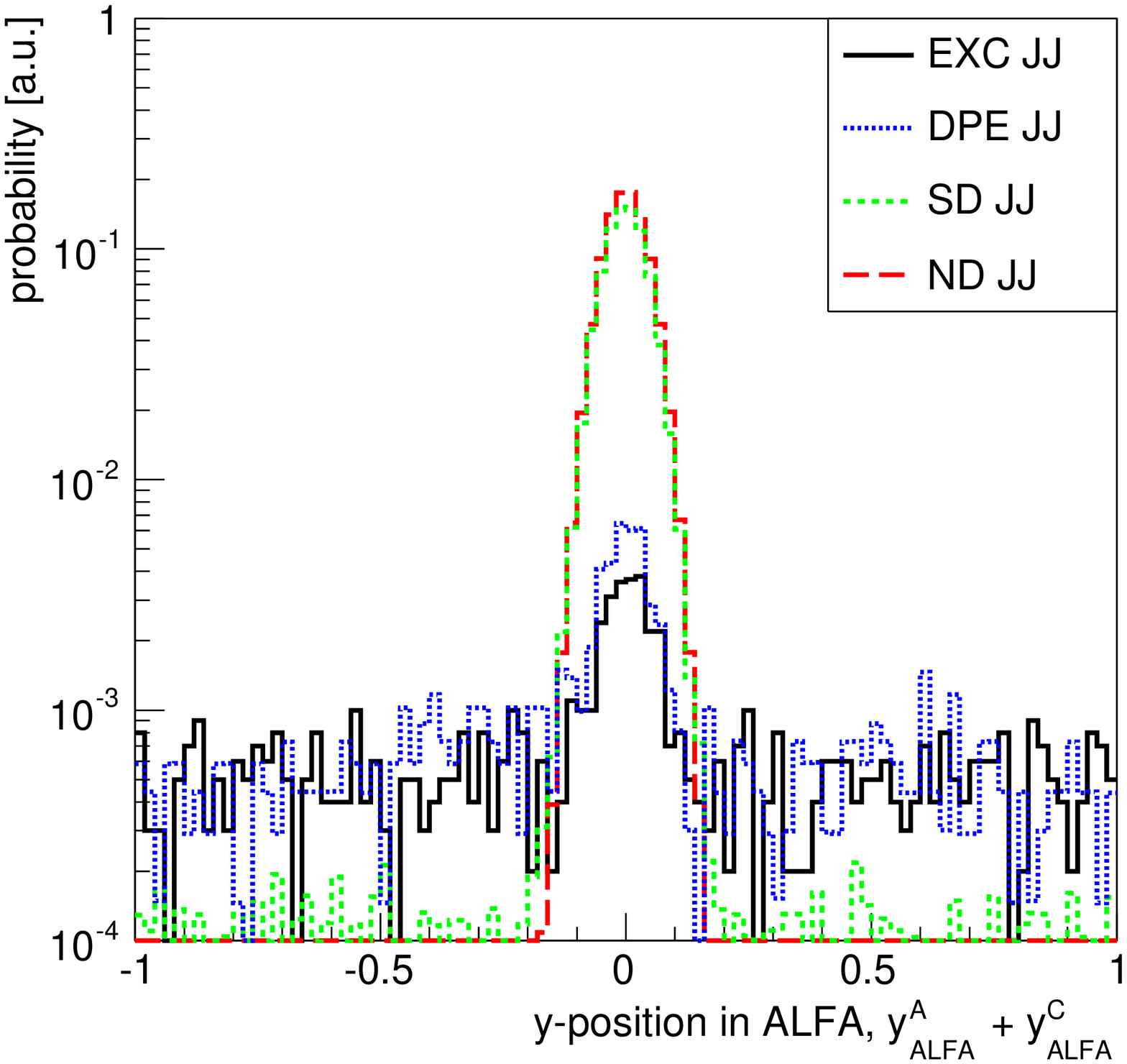}\hfill\hfill
  \vspace{-0.3cm}
\caption{Difference between y-position in ALFA stations. The peak around zero is mainly due to the jets produced together with protons scattered elastically.}
\label{fig_exc_jj_elastic_cut}
\end{figure}

The $\beta^* = 90$~m optics was designed to enhance the acceptance for the elastic scattering. When produced together with non-exclusive jets, such events will contribute as a background and need to be removed. This can be done using the properties of the elastic scattering. Transverse momenta of protons scattered elastically is expected to have exactly the same value but an opposite direction. In the ALFA detector such protons are expected to have top-bottom and left-right symmetry. Therefore, as is shown in Fig.~\ref{fig_exc_jj_elastic_cut}, the background can be efficiently removed by requiring $|y_A^{ALFA} + y_C^{ALFA}| < 1.5$ mm, where $A$ and $C$ stands for left or right side of the IP.

For further background reduction exactly one vertex reconstructed in the central detector was required. Next, the energy of scattered proton was correlated with the one from reconstructed jet system. Finally, exclusivity criteria such as the number of tracks reconstructed outside the jet system and the energy detected in the forward calorimeter were considered. All above requirements were applied similarly to the procedure described in~\cite{EXC_JJ_one_tag}.

\subsection{Results}
\begin{figure}
  \centering
  \includegraphics[width=0.49\columnwidth]{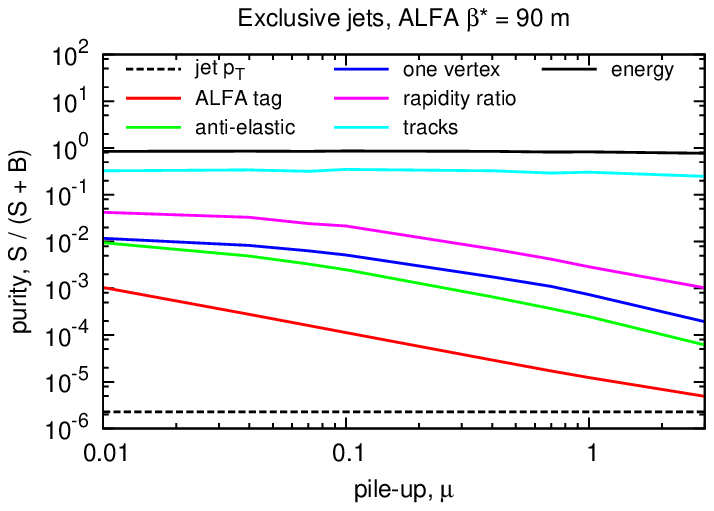}\hfill
  \includegraphics[width=0.49\columnwidth]{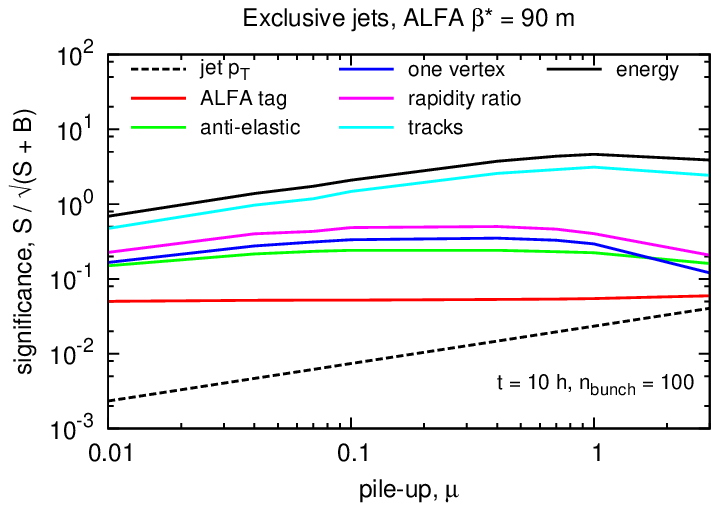}
  \vspace{-0.3cm}
\caption{Purity (left) and statistical significance (right) after a given selection as a function of pile-up.}
\label{fig_exc_jj_results}
\end{figure}

The increase of signal purity, defined as ratio of signal (S) to sum of signal and background (S + B) events, after the selection is shown in Fig. \ref{fig_exc_jj_results} (left). Pure ($>90\%$) signal is expected for all considered pile-up conditions ($0.01 < \mu < 2$). The quality of the measurement is expressed in terms of the statistical significance, $\frac{S}{\sqrt{S+B}}$. Its distribution after given selection cut as a function of pile-up is presented in Fig. \ref{fig_exc_jj_results}. The maximal significance is obtained for pile-up of about 1. A slow decrease for $\mu < 1$ is due to the amount of data that could be collected during the fixed time, whereas the decrease for $\mu > 1$ is a consequence of the single vertex requirement.

\section{Summary}
Measurement of double tagged high $p_T$ exclusive jets will be possible using the ATLAS/AFP detectors. It will be very challenging, but after the event selection, a signal-to-background ratio of 0.57 (0.16) for $\mu$ = 23 (46) can be achieved.

The semi-exclusive measurement in which only one proton is tagged was performed for four data-taking scenarios: AFP and ALFA detectors and $\beta^* = 0.55$ m, $\beta^* = 90$ m optics. After the signal selection, the signal-to-background ratio was found to between 5 and $10^4$, depending on the considered run scenario. The significant measurement can be carried out for the data collection period of about 100 h with pile-up of about 1.

Finally, a double tagged exclusive measurement of low-$p_T$ jets was shown to be feasible using $\beta^* = 90$ m optics and ALFA detectors. Pure ($>90\%$) and statistically significant signal is expected for all considered pile-up conditions ($0.01 < \mu < 2$).

\end{document}